\documentclass{aa}

\usepackage{txfonts}
\usepackage{hyperref}
\usepackage{graphicx}	
\usepackage{amsmath}	
\usepackage{amssymb}	
\usepackage{xspace}
\usepackage{subcaption} %

\defcitealias{Maiolino25}{M25}
\defcitealias{Ignas25}{J25}
\defcitealias{Scholtz23}{S25}

\newcommand{\cmsq}{$\mathrm{cm^{-2}}$\xspace}
\newcommand{\chandra}{\textit{Chandra}\xspace}

\newcommand{\kbol}{\mbox{$k_{\mathrm{bol},X}$}}

\begin{document} 
\title{JWST-discovered AGN: evidence for heavy obscuration in the type-2 sample from the first stacked X-ray detection}
\titlerunning{Evidence for heavy X-ray obscuration in type-2 AGN discovered by JWST}
\authorrunning{A. Comastri et al.}
   \author{A. Comastri\thanks{andrea.comastri@inaf.it}\inst{1} \and
   	          G. Lanzuisi\inst{1} \and
   	          F. Vito\inst{1} \and
   	          S. Marchesi\inst{2,1,3} \and
   	          M. Brusa \inst{2,1} \and 
                    R. Gilli \inst{1} \and 
                    I. Juodžbalis \inst{4,5}
                    R. Maiolino \inst{4,5,6}
                    G. Mazzolari\inst{7,1}
                    G. Risaliti \inst{8,9} \and
                    J. Scholtz \inst{4,5} \and
                    C. Vignali \inst{2,1} 
          }
 \institute{
INAF -- Osservatorio di Astrofisica e Scienza dello Spazio di Bologna (OAS), Via Gobetti 93/3, I-40129 Bologna, Italy
  \and
Dipartimento di Fisica e Astronomia (DIFA), Università di Bologna, via Gobetti 93/2, I-40129 Bologna, Italy 
\and
Department of Physics and Astronomy, Clemson University, Kinard Lab of Physics, Clemson, SC 29634, USA \and
Kavli Institute for Cosmology, University of Cambridge, Madingley Road, Cambridge CB3 OHA, UK \and
Cavendish Laboratory – Astrophysics Group, University of Cambridge, 19 JJ Thomson Avenue, Cambridge CB3 OHE, UK \and
Department of Physics and Astronomy, University College London, Gower Street, London WC1E 6BT, UK \and
Max-Planck-Institut für extraterrestrische Physik (MPE), Gießenbachstraße 1, 85748 Garching, Germany \and
Dipartimento di Fisica e Astronomia, Università di Firenze, via G. Sansone 1, I-50019 Sesto Fiorentino, Firenze, Italy \and
INAF -- Osservatorio Astrofisico di Arcteri, Largo Enrico Fermi 5, I-50125 Firenze, Italy 
  }

\date{}
\abstract{
One of the most puzzling properties of the high-redshift AGN population recently discovered by JWST, including both broad-line and narrow-line sources, is their X-ray weakness. With very few exceptions, and regardless of the optical classification, they are undetected at the limits of the deepest Chandra fields, even when stacking signals from tens of sources in standard observed-frame energy intervals (soft, hard, and full bands). It has been proposed that their elusive nature in the X-ray band is due to heavy absorption by dust-free gas or intrinsic weakness, possibly due to high, super-Eddington accretion.
In this work, we perform X-ray stacking in three customized rest-frame energy ranges (1-4, 4-7.25, and 10-30 keV) of a sample of 50 Type 1 and 38 Type 2 AGN identified by JWST in the CDFS and CDFN fields. For the Type 2 sub-sample, we reach a total of about 210 Ms exposure, and we report a significant ($\sim 3\sigma$) detection in the hardest (10--30 keV rest frame) band, along with relatively tight upper limits in the rest frame softer energy bands. The most straightforward interpretation is in terms of heavy obscuration due to gas column densities well within the Compton thick regime ($> 2 \times 10^{24} $cm$^{-2}$)  with a large covering factor, approaching 4$\pi$. 
The same procedure applied to the Type 1 sub-sample returns no evidence for a significant signal in about 140 Ms stacked data in any of the adopted bands, confirming their surprisingly elusive nature in the X-ray band obtained with previous stacking experiments. 
A brief comparison with the current observations and the implications for the evolution of AGN are discussed.
 } 


\maketitle
%
\section{Introduction}\label{intro}

Several surveys with the James Webb Space Telescope (JWST) have discovered a population of high redshift (z $\sim$ 4--10) Active Galactic
Nuclei (AGN) with bolometric luminosities significantly
lower ($<10^{45}$ erg s$^{-1}$) than previously observed in bright quasars dominating the AGN population at high redshift. 
Most of them have been identified via detection of relatively broad emission lines with H$\alpha$ widths of the order of 2000 km s$^{-1}$ and are thus classified as Type 1 AGN (e.g., \citealt{Maiolino23}, \citealt{Ignas25}, hereafter \citetalias{Ignas25}). Their black-hole mass estimates range from 10$^6$ to 10$^8$ M$_{\odot}$ \citep[e.g.,][]{Maiolino24, Kocevski23}.They tend to be over-massive compared to their host galaxies’ stellar masses, and with respect to the local relation, suggesting a vigorous growth phase at early times (although this interpretation is still debated; see \citealt{Tonina24}). About 10\%--30\% of them \citep{Hainline2025} are dubbed
 Little Red Dots owing to their point-like appearance and red near-infrared colors; in addition to broad permitted lines, they are characterized by a V-shaped continuum, with a relatively blue UV slope \citep[e.g.][]{Kocevski25}.  
Additionally, a growing number of narrow-line Type 2 AGN are identified using several diagnostic diagrams for the emission line, some specifically developed for the classification of the emission lines at high redshift \citep{Scholtz23,Mazzolari24b,Calabro23}.
The space densities of these newly,  JWST-discovered AGN at z $\sim$ 4--6 is about 1--2 order of magnitude higher than the extrapolation of the QSO luminosity function and of the X-ray selected AGN luminosity function \citep[e.g.,][]{Harikane23,Matthee24,Greene24,Maiolino2024_JADES,Ignas25} suggesting that JWST is tracing the emergency of a new population, much larger than the most optimistic \citep{Giallongo15} estimates from the X--ray selected AGN. Given that known X-ray emitting AGN accounts for some 80- 90\% of the X-ray background \citep[e.g.,][]{Moretti03,Nico17}, it is not surprising that this new population is X-ray silent.
Although many examples of X--ray weak quasars were reported in the literature \citep{Risaliti2001} well in advance the JWST surveys, the JWST population is much larger. Even more intriguing, this new population remains undetected even with stacking techniques performed in deep X--ray observations such as those in the Chandra Deep Fields  (\citealt{Yue24}, \citealt{Mazzolari24}, \citealt[hereafter, \citetalias{Maiolino25}]{Maiolino25}).

Two broad classes of models were proposed to explain the X-ray weakness:
Super-critical accretion resulting in an extremely soft X--ray
spectrum \citep[e.g.,][]{Pacucci24,MadauHaardt24,Madau25,Maiolino25} and significant absorption at high inclinations due to self-shadowing, or 
heavy obscuration with a high covering factor and column densities exceeding
$10^{24}\,\mathrm{cm^{-2}}$ by dust free/poor gas to account for the
presence of broad lines and UV emission (e.g., \citetalias{Maiolino25}, \citealt{Ji25}, \citealt{Ignas24}).

In the former case, the steep X-ray emission explains the lack of
detection in the observed standard (0.5-2 and 2-7 keV) bands
corresponding, approximately, to 4 to 40 keV rest frame.  The heavy absorption hypothesis implies that the X-ray flux is dimmed
by obscuration, especially in the soft band.
For optical depths of the order of $\tau >> 1$ and large covering
factors, the signal in the hard X--ray band may also be strongly suppressed.

In order to understand which physical mechanism is more likely
to explain the observed properties, we carried out a \chandra stacking analysis at the positions of the
sources in our sample in several energy intervals, tuned to detect
characteristic features of heavy absorption, such as a very hard
X--ray spectrum emerging at $>$ 8 keV, enhanced emission around the
iron line complex at $\sim$ 6--7 keV and/or the presence of a strong soft
excess below about 1 keV.  

A detection in the stacked signal in the softest energy
channels would be suggestive of a steep spectrum, possibly related to
supercritical accretion, while a signal at the
hardest energies would favor the heavy obscuration interpretation.
In principle, one may constrain the average column densities owing to the
strong dependence of the X--ray flux upon the optical depth for
Compton scattering.
The present work takes a step further with respect to previous analyses based on the search for a stacked signal in deep \chandra fields as we explicitly stack emission from all sources in the samples in the same rest-frame bands, and fully exploit the stacked X-ray SED shapes to estimate the absorption levels.

The paper is organized as follows. In Section 2, the sample selection is presented, in Section 3 we describe the procedure adopted to analyze (stack) the X-ray data; in Section 4, we present our results; in Section 5, we discuss the implications of our results concerning the current models of the X-ray and multi-band properties of JWST AGN and draw conclusions. 
Errors are reported at the 68\% confidence levels, while limits are given
at the 90\% confidence levels. 
We adopt a flat cosmology with $H_0=67.7\,\mathrm{km\,s^{-1}}$ and $\Omega_m=0.307$ \citep{Planck16}.

\section{The JWST sample of AGN}\label{sample}

Our parent sample is composed of AGN identified by JWST in the footprints of GOODS-N and GOODS-S fields, the deepest \chandra fields, with a total exposure of about 2 and 7 million seconds, respectively. The choice is dictated by the need for deep X-ray exposures to obtain meaningful limits on the AGN X-ray emission.
The optical classification as Type 1 and Type 2 AGN is mainly from the JWST Advanced Deep Extragalactic Survey (JADES; \citealt{Eise2023}, \citealt{DEugenio2025JADES}). 
Type 1 AGN have broad Balmer emission lines with Full Width Half Maximum (FWHM) $>$ 1000 km s$^{-1}$ \citep{Matthee24} .
The Type 2 sample is originally published in \cite{Scholtz23} (hereafter \citetalias{Scholtz23}) and reported in \citetalias{Maiolino25}. The Type 1 objects are published in \citetalias{Ignas25} and also in \citetalias{Maiolino25}. We refer to the discovery papers for details on the selection and classification. The list of sources is reported in the Appendix in Tables~\ref{Tab_phy_prop_type1} (Type 1) and \ref{Tab_phy_prop_type2} (Type 2). We only used the 'robust' AGN candidate in the CDFS from \citetalias{Ignas25} and removed the 'tentative' candidates from our sample. 

The sources' X-ray emission is discussed in \citetalias{Maiolino25} and \citetalias{Ignas25}. A few sources were detected in the X-ray band and excluded from our sample.  Two broad-line AGN (GN 721 and XID\_403\footnote{The ID refers to the \textit{Chandra} 4 Ms catalog \citep{Xue11}.}) are relatively faint in the X-rays ($\kbol>100$, where $\kbol=L_{bol}/L_{2-10keV}$), but with a hard X-ray spectrum consistent with Compton thick obscuration \citep{Gilli14,Circosta19}. The absorption-corrected bolometric correction derived from Compton thick obscuration is consistent with the observed distribution of standard Type 1 AGN. 
Other two broad-line AGN (GS\_49729 and GS\_209777) are detected with an unobscured X-ray spectrum ($\Gamma\sim1.7$) and average bolometric correction $\kbol\sim15$, hence also consistent with standard type 1 AGN.
One Type 2 object (GS\_21150) is X-ray detected, but with a number of photon counts insufficient to estimate the spectral shape.

 In the following, we consider the 38 Type 2 and 50 Type 1 AGN that are not individually detected, and that do not 
produce a significant stacked signal in the standard, observed-frame soft (0.5--2 keV) and hard (2--7 keV) \chandra bands (\citetalias{Maiolino25}).

\begin{figure*}
\centering
\begin{subfigure}[t]{0.48\textwidth}
    \centering
    \Large \textbf{Type 2s}\par\vspace{0.1cm}
    \includegraphics[width=\textwidth,keepaspectratio]{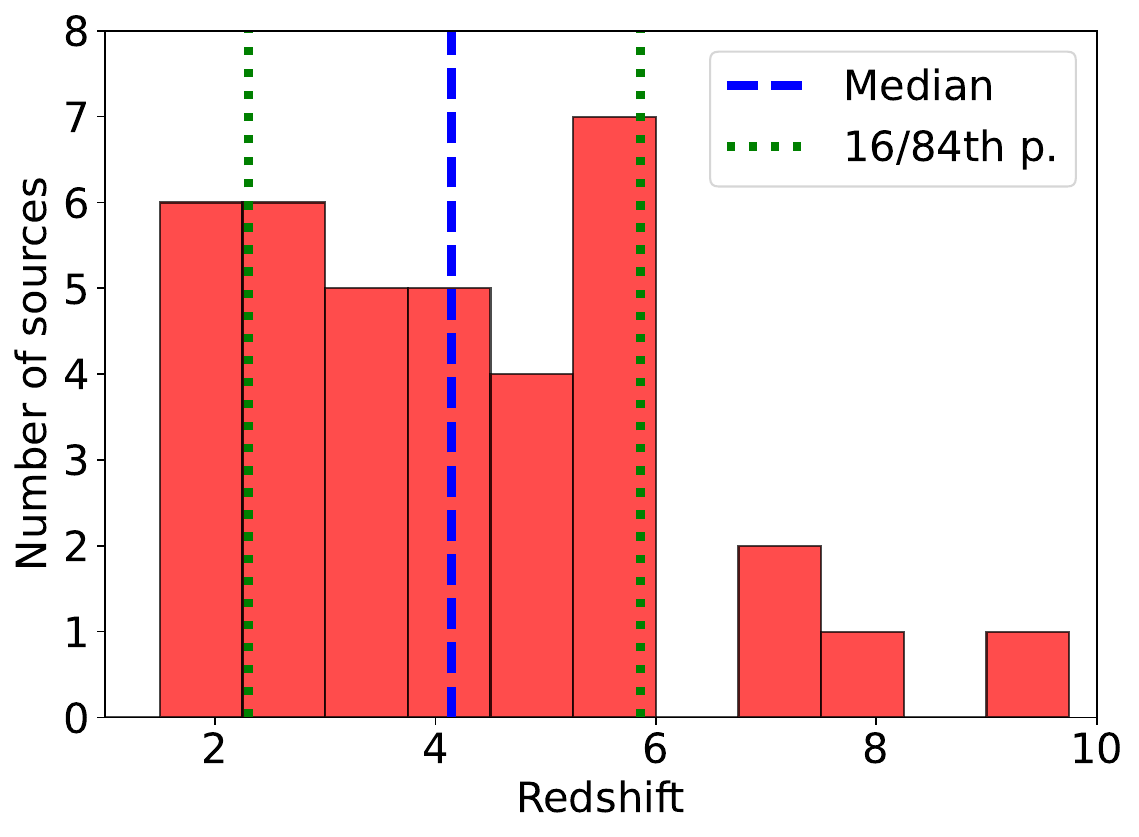}
\end{subfigure}
\hfill
\begin{subfigure}[t]{0.48\textwidth}
    \centering
    \Large \textbf{Type 1s}\par\vspace{0.1cm}
    \includegraphics[width=\textwidth,keepaspectratio]{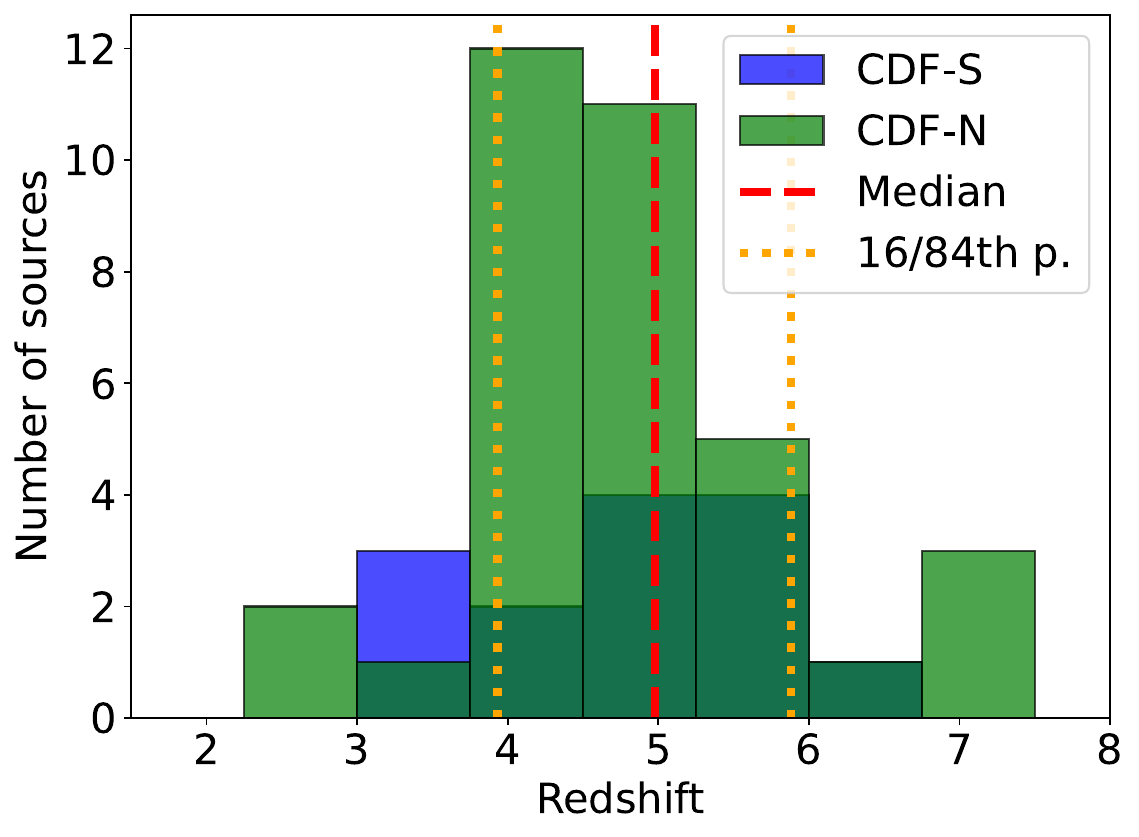}
\end{subfigure}
	\caption{Left: Redshift distribution for the Type 2s AGN analysed in this work, all from CDF-S. The median of the distribution is reported, as well as the 16th and 84th percentiles, as labeled. 
    Right: Redshift distribution for the Type 2s AGN analysed in this work, from the CDF-S (blue) and CDF-N (green). The median of the distribution is reported, as well as the 16th and 84th percentiles, as labeled. 
	}\label{fig:redshift_distribution}
\end{figure*}

The relevant data for the two samples are reported in Appendix~\ref{sec:appendixA}, in Tables~\ref{Tab_phy_prop_type1} and \ref{Tab_phy_prop_type2}. 
The redshift distribution for both samples is reported in the two panels of Figure~\ref{fig:redshift_distribution}. As a consequence of the JADES parent sample's selection criteria, most sources are at $z >$ 2, extending up to $z\sim$9. The median redshifts are $z\sim4.2$ for Type 2 and $z\sim5$ for Type 1. 

For type 2 AGN, the bolometric luminosity is inferred from the extinction-corrected [O III]5007 line, adopting the scaling relation given in \citetalias{Scholtz23}, and is reported in Table~\ref{Tab_phy_prop_type2} and Figure~\ref{fig:lbol_lx}. As can be seen in the figure, the bolometric luminosities span more than three orders of magnitude in the range log$\frac{L_{bol}}{\mathrm{erg\,s^{-1}}}\sim $41.5--45.0.  

The 38 Type-2s from \citetalias{Scholtz23} all lie in the CDFS, within 3.3 arcmin from the field center. Thus, we take advantage of the sharpest \chandra psf (90\% encircled counts fraction always within 2.5 arcsec radius).
Type-1s in the CDFN (36 sources) lie in the central region of the field, within 3-4 arcmin from the aimpoint of the majority of the pointings contributing to the total 2Ms exposure. 
Half of the Type-1s in the CDFS (7 sources) lie in the central 3 arcmin region of the field, while the other half are selected from different JWST pointings and are located between 4 and 7 arcmin (90\% encircled counts fraction goes from 3 to 6 arcsec radius).

\section{Data reduction and stacking procedure}\label{data_reduction}

To increase the sensitivity of the X-ray stacking analysis, we performed a stack in 3 adaptively chosen bands to test the hypothesis of heavy obscuration in the average spectrum of the sources in the sample.

The X-ray data in the \chandra deep fields were reduced and extensively analyzed by many authors. For the present paper, we rely on the data products (e.g., stacked event files, exposure maps, source catalogs) made available by the Penn State \chandra team \footnote{available at \url{https://personal.science.psu.edu/wnb3}}. We refer to \cite{Luo17} and \cite{Xue16} for all the details of the \chandra data analysis.




\subsection{Stacking procedure}\label{Xray_reduction}

In order to obtain a robust estimate of the background, the detected sources from the \cite{Luo17} catalog were removed from the event files, assuming circular regions whose radius is a function of the source intensity: from 4 pixels ($2^{\prime\prime}$)
for $<30$ full band counts to 6 pixels ($3^{\prime\prime}$) for bright $>100$ counts sources. The adopted radii exclude $95-98\%$ of the counts of the detected sources. This ensures $<1$ count per source on average, leaking outside the excluded region. 
The local background was estimated in a box of 20 pixels on each side around the excluded regions. The masked pixels were then assigned random discrete values extracted from a Poisson distribution with expectation value set to the median background counts per pixel, thus reproducing the same background levels. For each source to be stacked, square cut-outs of 50 pixels were created around the input position. 

The novelty of our approach with respect to previous works that stacked Chandra images at the position of known JWST sources lies in the careful selection of the energy bands where the stacking is performed. Specifically, we selected three rest-frame bands (1--4, 4--7.25, and 10--30\,keV), corresponding to different observed bands depending on the redshift of each source included in the stacking. In the rest of the paper, we will refer to these bands as soft band (SB, 1--4 keV), medium band (MB, 4--7.5 keV), and ultra hard band (UHB, 10--30 keV). 
The 1--4\,keV band was chosen to include the softest photons detectable by Chandra in the observed energy range, while we chose the 10--30\,keV band as our hardest band after performing multiple tests\footnote{The other bands we tested are the 8--20\,keV, 8--30\,keV, and 10--20\,keV ones} that allowed us to determine that such an energy range is the best trade-off between the band width, to maximize the total counts, and the high energy boundary to minimize the background counts.  We limit the MB to 7.25 keV, a trade-off between the band width and the expected spectral shape, which drops quickly, especially for obscured sources, above the Iron K-edge.

The total stacked counts (Table~\ref{Tab_Xray_cts}) are extracted from a circular region centered at the optical JWST position with $R=2^{\prime\prime}$,  whereas the background counts are measured from an annular region with inner radius of $5^{\prime\prime}$ and outer radius of $10^{\prime\prime}$ and then rescaled for the ratio background/source area. The net source counts are then computed by subtracting the rescaled background counts from the total counts. The errors are propagated accordingly.


\begin{figure*}
	\begin{center}	
        \textbf{\Large Type 2 stack}\\[0.5em] 
\includegraphics[width=0.3\textwidth,keepaspectratio]{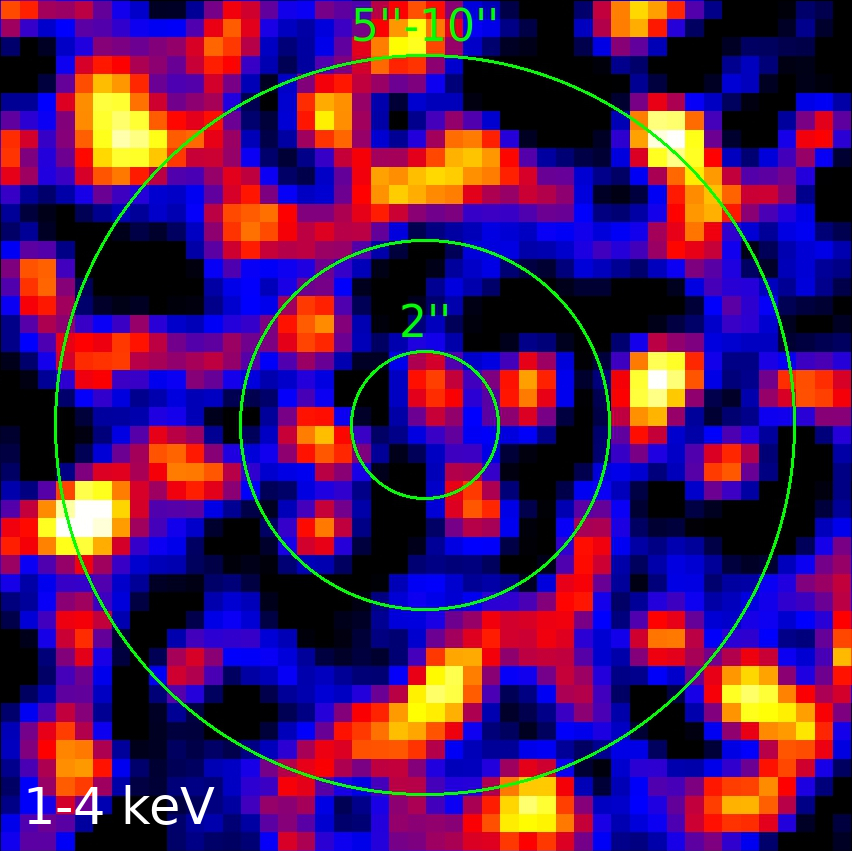} 
    \vspace{0.1cm}    \includegraphics[width=0.3\textwidth,keepaspectratio]{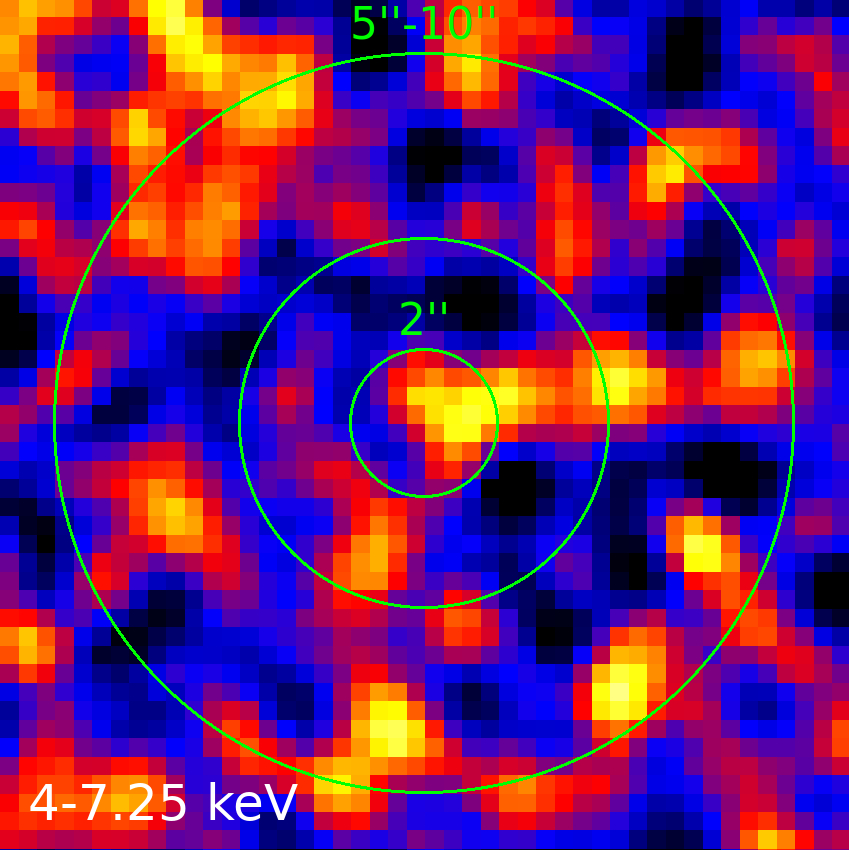}
    \vspace{0.1cm}    \includegraphics[width=0.3\textwidth,keepaspectratio]{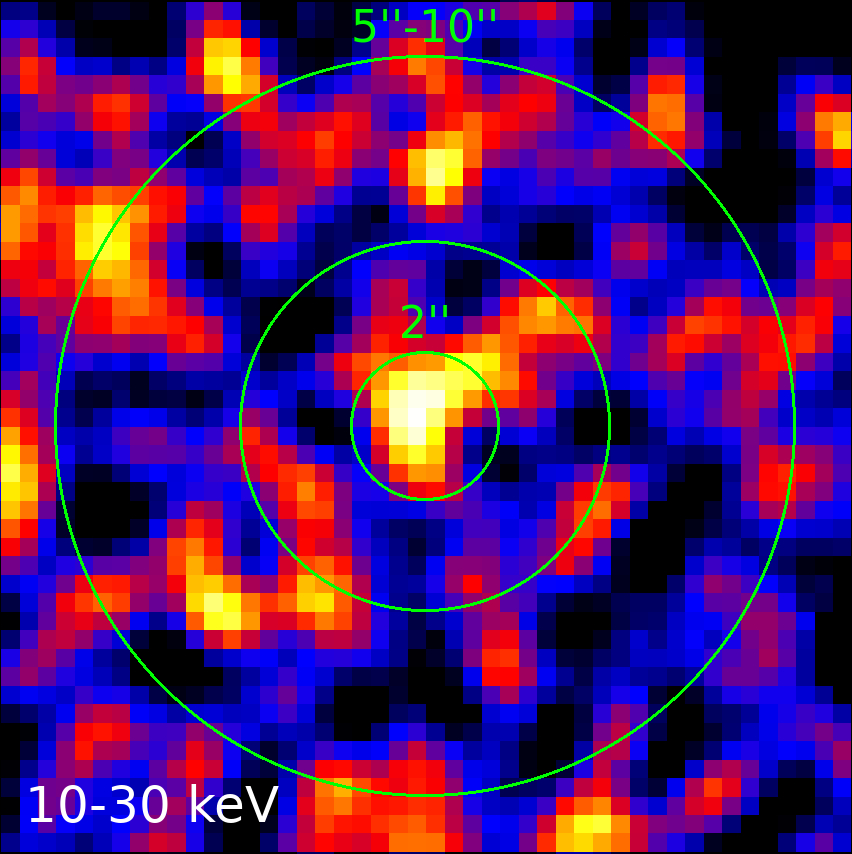}
    
	\end{center}
	\caption{Left: Type 2 stacked image 1-4 keV rest-frame. 
    Center: Type 2 stacked image 4-7.25 keV rest-frame.
    Right: Type 2 stacked image 10-30 keV rest-frame.
	}\label{fig:stack_type2}
\end{figure*}

\begin{table*}
	\centering
	\caption{Stacking results for the Type 1 and Type 2 samples. Total and Background counts in the SB, MB and UHB. The source counts are net counts. The background counts are scaled to the size of the source region.}
	\begin{tabular}{cccccc} 
		\hline
		\multicolumn{1}{c}{{ Sample}} &
		\multicolumn{1}{c}{{ $T_{exp}$ }} & & 
		\multicolumn{3}{c}{{ Total Counts}} 	\\ 
		\cline{4-6}
		\multicolumn{1}{c}{{}} &			
		\multicolumn{1}{c}{{ [Ms]}} &  & 
		\multicolumn{1}{c}{SB} &
		\multicolumn{1}{c}{{MB }} &
		\multicolumn{1}{c}{{ UHB}} 			\\

          Type 1  & 139.8 & SRC & $<11.4$         & $<12.2$       & $<34.6$        \\
                  &       & BKG & $ 125.6\pm2.6$  & $148.1\pm2.8$ & $639.5\pm5.9$  \\   
		 Type 2  & 209.4 & SRC & $<16.6$         & $<41.7$ & $122.8\pm38.3$ \\
                  &       & BKG & $263.8\pm3.8$   & $295.5\pm4.0$ & $1272.2\pm8.3$ \\  
		\hline
	\end{tabular} \label{Tab_Xray_cts}\\
	\tablefoot{The Type 1 exposure i 59.6Ms for CDFN and 80.2Ms for CDFS. 
}
\end{table*}

\begin{figure*}
	\begin{center}		
        \textbf{\Large Type 1 stack}\\[0.5em] \includegraphics[width=0.3\textwidth,keepaspectratio]{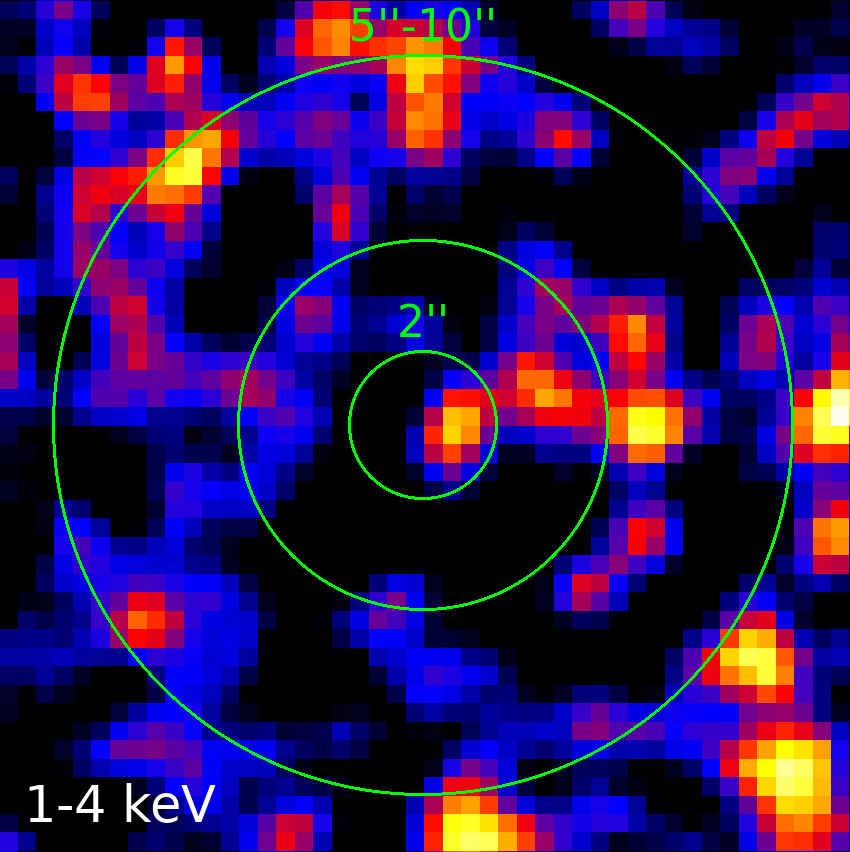} 
    \vspace{0.1cm}    \includegraphics[width=0.3\textwidth,keepaspectratio]{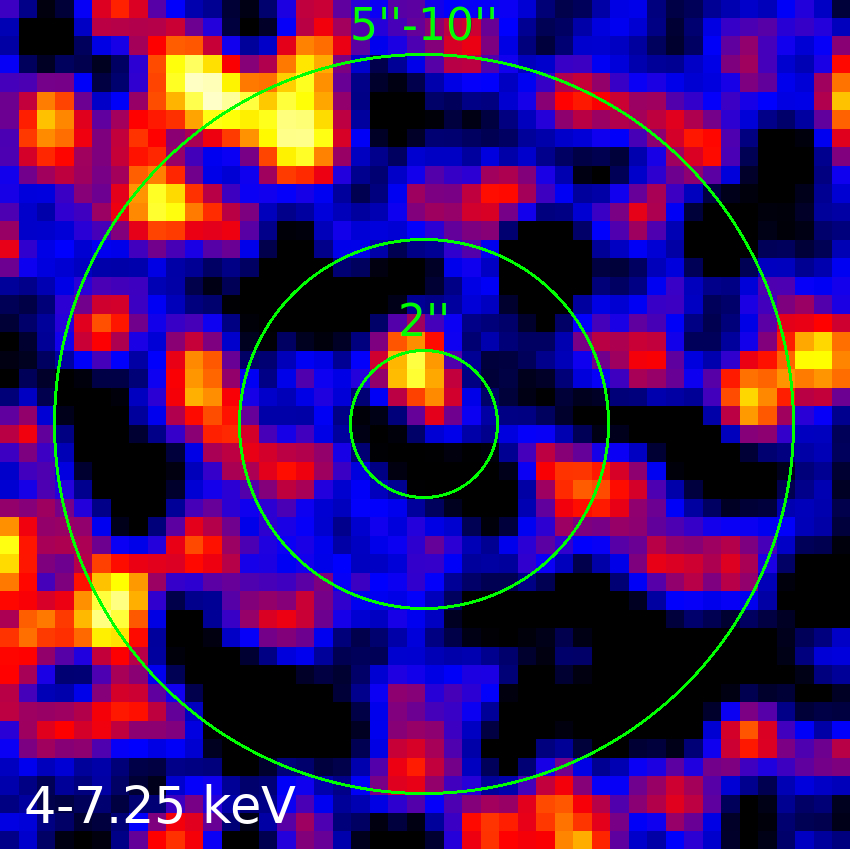}
    \vspace{0.1cm}    \includegraphics[width=0.3\textwidth,keepaspectratio]{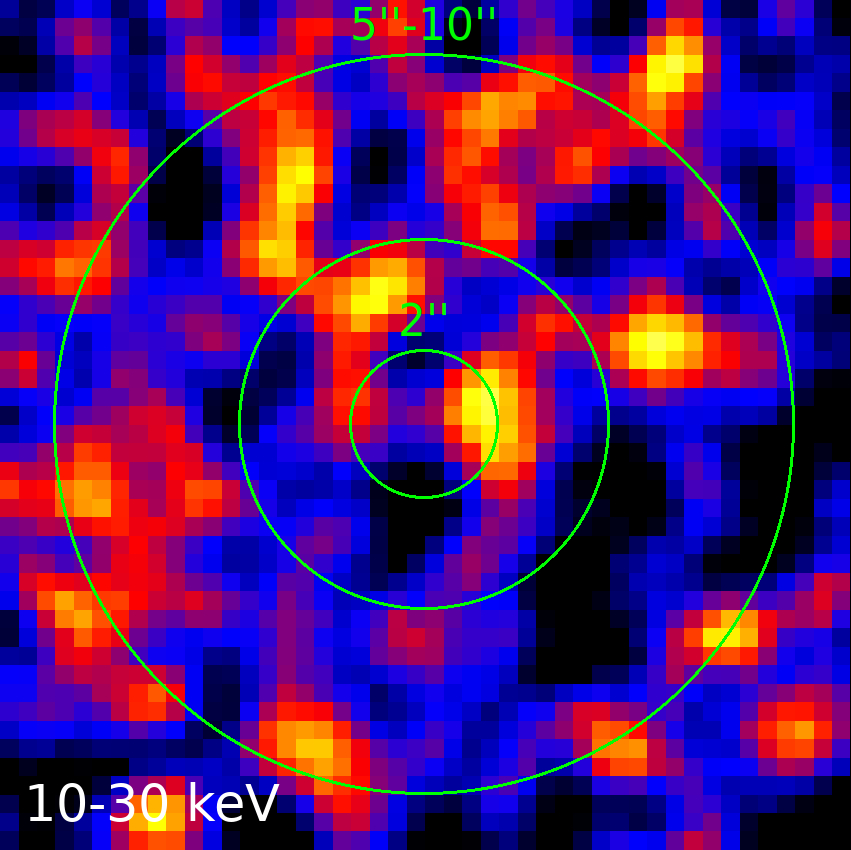}
    
	\end{center}
	\caption{Total (CDFS+CDFN) stacked images for the Type 1 sample. Left:  1-4 keV SB. 
    Center: 4-7.25 keV MB. Right: 10-30 keV UHB.
	}\label{fig:stack_type1}
\end{figure*}

\begin{figure}
	\begin{center} \includegraphics[width=0.45\textwidth,keepaspectratio]{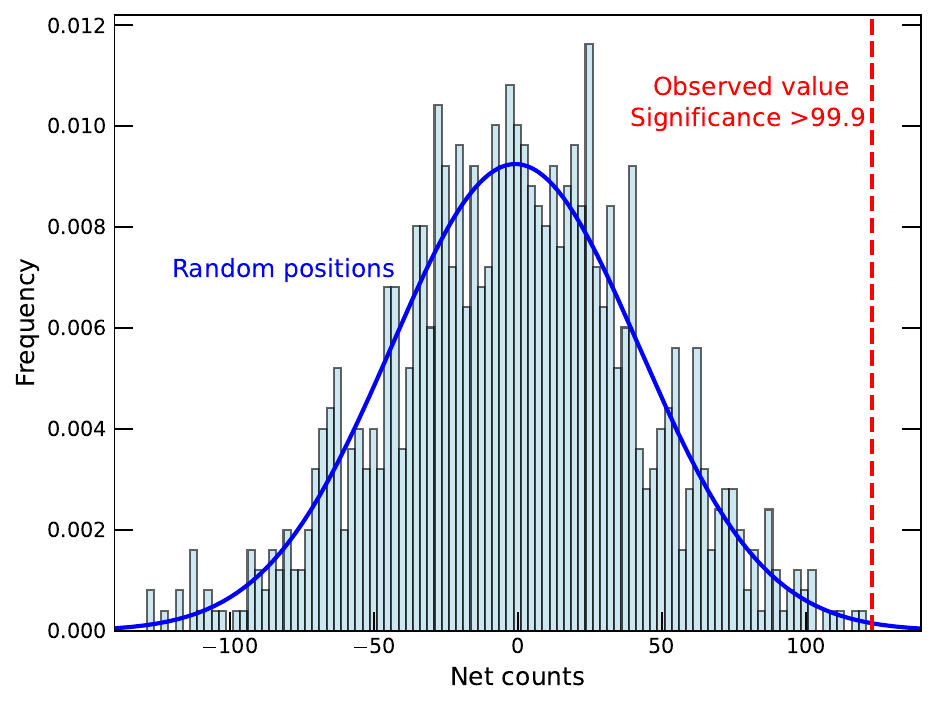} 
	\end{center}
	\caption{Distribution of stacked net-counts in the UHB for 1000 realizations of random positions (cyan histogram). The number of stacked net counts obtained for the type-2 AGN sample is marked with the dashed red line. All of the random realizations return a lower number of net counts, implying a significance of $>0.999$ of the stacked emission from the type-2 AGN sample.
}\label{fig:simulations}
\end{figure}

\begin{figure}
	\begin{center} \includegraphics[width=0.5\textwidth,keepaspectratio]{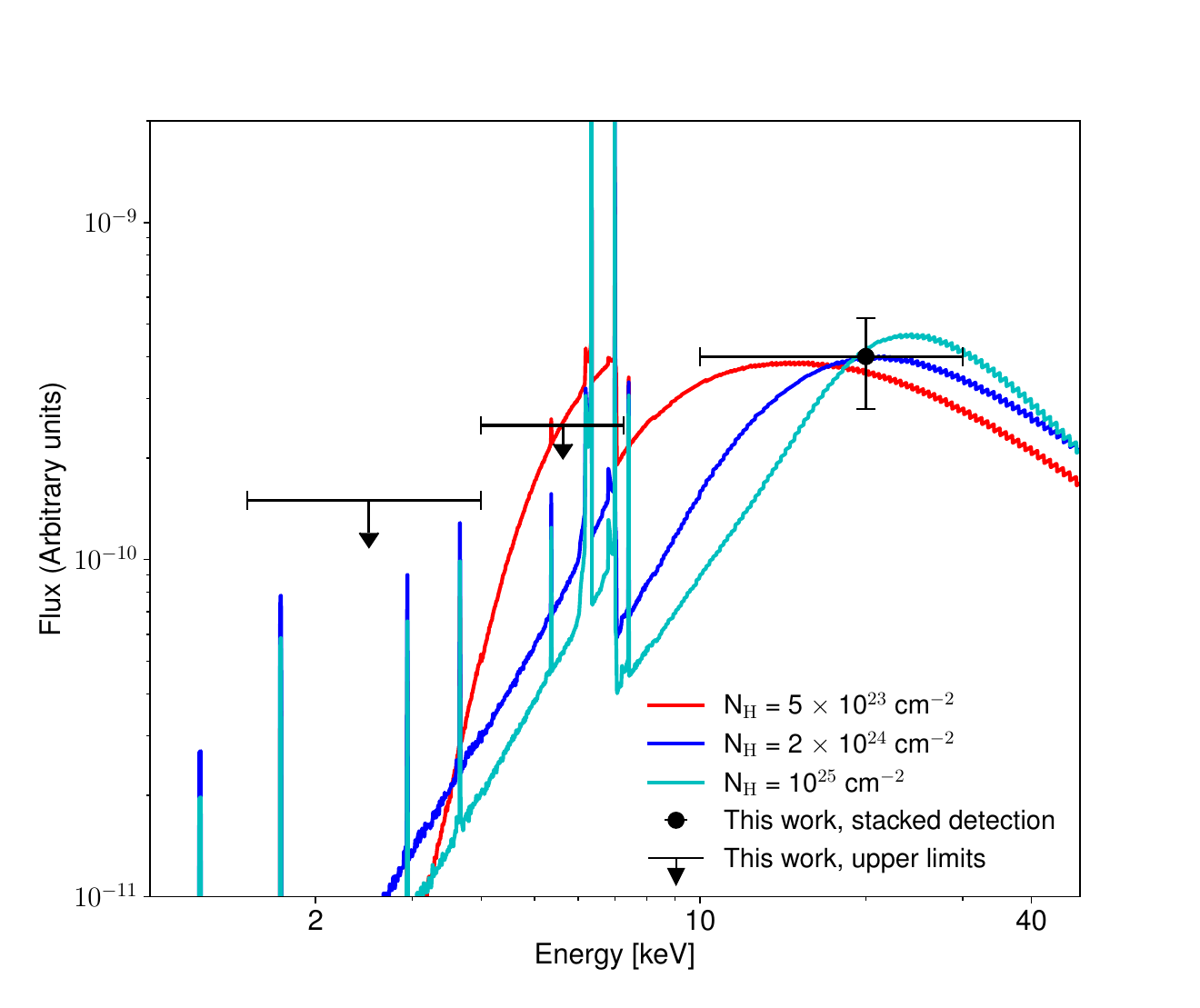} 
	\end{center}
	\caption{Spectral models computed with the {\it uxclumpy} code described in \citep{Buchner19} rescaled to the observed 10--30 keV flux from the stacking analysis. We refer to the text for details on the model assumptions and geometry. The column density increases from 5 $\times$ 10$^{23}$ cm$^{-2}$ (red curve), to 2 $\times$ 10$^{24}$ cm$^{-2}$ (blue curve) and 10$^{25}$ cm$^{-2}$ (cyan curve).  The UHB detection and the upper limits in the SB and MB are also reported.
}\label{fig:spectra_uxclumpy_w_UL}
\end{figure}

\section{Results}\label{results}

\begin{table*}
	\centering
	\caption{Flux and Luminosity limits}
	\begin{tabular}{ccccccc} 
		\hline
		\multicolumn{1}{c}{{ Sample}} &
            \multicolumn{1}{c}{{ Field}} &
		\multicolumn{1}{c}{{ $T_{exp}$ }} & & 
		\multicolumn{3}{c}{{ Bands}} 	\\ 
		\cline{5-7}
		\multicolumn{1}{c}{{}} & 	
            \multicolumn{1}{c}{{}} & 			
		\multicolumn{1}{c}{{ [Ms]}} &  & 
		\multicolumn{1}{c}{SB} &
		\multicolumn{1}{c}{{MB }} &
		\multicolumn{1}{c}{{ UHB}} 			\\

          Type 1  & CDFS  & 80.3  & Flux  & $<2.5\times10^{-18}$ & $<6.7\times10^{-19}$ & $<5.2\times10^{-18}$ \\
                  &       &       & Lum   & $<6.1\times10^{41}$ & $<1.6\times10^{41}$ & $<1.3\times10^{42}$ \\
                  & CDFN  & 60.0 & Flux  & $<1.3\times10^{-18}$ & $6.6\times10^{-19}$ & $<1.1\times10^{-17}$ \\
                  &       &       & Lum   & $<3.7\times10^{41}$ & $<1.9\times10^{41}$ & $<3.2\times10^{42}$ \\
 		\hline \\
		 Type 2  & CDFS &  209.4 & Flux & $<7.3\times10^{-19}$    & $<1.1\times10^{-18}$ & $1.15\pm0.36\times10^{-17}$ \\
                  &      &        & Lum  & $<1.4\times10^{41}$ & $<2.1\times10^{41}$ & $2.14\pm0.66\times10^{42}$ \\  
		\hline
	\end{tabular} \label{Tab_Xray_flux}\\
	\tablefoot{...
}
\end{table*}

The thumbnails associated with the rest-frame stack in the SB, MB and UHB 
bands are shown in Figure~\ref{fig:stack_type2} and Figure~\ref{fig:stack_type1} for the Type 2 and Type 1 AGN, respectively. The background and source counts for both samples, are reported in Table~\ref{Tab_Xray_cts}.

The AGN in the Type 1 sample were not detected in any rest-frame bands. The sources in the Type 2 AGN sample were not detected in the SB and MB (as defined above), while they were detected in the UHB, with significance $(1-P_B)=0.9996$ using Poisson statistics (0.9993 with Gaussian statistics).
We also considered the possibility that the UHB signal is due to a specific subset of sources. Several tests were performed by dividing the Type 2 sample according to their redshift, bolometric luminosity, and "expected" X-ray flux, for a given bolometric luminosity and redshift, assuming the standard X-ray bolometric correction from \citep{Duras20}. The tests were performed using only two bins due to the low number of candidates. There is no evidence for a signal in any of the adopted bins, suggesting the lack of a trend with either redshift or luminosity. 

The significance of the detection in the UHB was also tested via simulations: we generated 1000 realizations of the Type 2 sample by randomizing the position of each source, in a region within one arcmin of the original position (to ensure exposure and background levels comparable with the real sources) and repeated the stack exercise. None of the realizations returns a number of counts higher than observed. 
The resulting detection significance is therefore $>0.999$, that is, $\gtrsim3\sigma$ (Figure \ref{fig:simulations}).

To convert stacked source counts into fluxes and then luminosities, we proceed as follows.
First, we assume three different spectral models for the intrinsic emission, either a power law with $\Gamma=2$ or a Compton-Thick obscured model using the {\tt uxclumpy} code \citep{Buchner19} with either $N_H=2\times10^{24}$ or $N_H=10^{25}$ \cmsq (see below for details on the model parameters).  
Then we use an average instrumental effective area obtained from a position in the same region as the stacked sources, both for CDF-S and CDF-N. The response files account for the variation of \chandra response over time since it is computed as an average of the 102 pointings in the CDF-S and the 20 pointings in the CDF-N collected over time in the two fields.
For each stacked object, we used its redshift to shift the intrinsic spectrum and convolve it with the effective area. This procedure allows us to obtain the expected distribution of detected photons in a given observed band corresponding to the SB, MB, and UHB rest-frame bands. 
We considered 1000 logarithmically spaced bins in the observed energy range 0.3-10 keV.
The average expected photon energy is computed for each source and each band as $\sum (E_i*N_i)/\sum (N_i)$, and is translated into a corresponding average effective area for that source. 
The sample-averaged photon energy and effective area, weighted for the effective area contributions of each source (in terms of exposure times, the contributions are roughly the same for each source), are then estimated. For example, Type 2 AGN in the UHB have average photon energies in the range $1.5-4.8$ keV, 
and average effective areas in the range $210-400$ \cmsq respectively, and the sample averages are 3.3 keV and 268 \cmsq.
The total exposure time is computed by stacking the values at the position of each source from the exposure maps of the two fields.
The fluxes are then computed, at the average observed band, as: $F=(Counts\times<E>)/(T_{exp} \times <A_{eff}>)$. 

At the same time, the luminosities are estimated assuming the corresponding average redshift, without k-correction, since the entire procedure is performed in the rest frame.
Since exposures and effective areas are very different between CDF-S and CDF-N, flux and luminosity limits in the CDFS and CDF-N were computed separately.

The differences between fluxes and luminosities computed with an unabsorbed power law spectrum and with a heavily absorbed one with $N_H=2\times10^{24}$ or $N_H=10^{25}$ \cmsq are of the order of $\sim20-30\%$ in the SB and minimal in the MB and UHB, while are negligible between the two CT cases; therefore, we report only the value for the $N_H=2\times10^{24}$ CT case in \ref{Tab_Xray_flux}.

We note that for the 1-4 keV band, a fraction of the rest-frame band moves below the \chandra low energy limit of 0.3 keV for sources at $z\gtrsim2$ (almost all our sources). When computing fluxes and luminosities, we must consider that only part of the intrinsic spectrum will be included in the observed band. 
We correct for this effect by computing the fraction of 'missing counts' for each source for a given spectral shape. 
For example, if 70\% of the intrinsic photons are expected to be missing below 0.3 keV, to recover the correct count rate, the exposure time is decreased by 70\%. This correction is larger for the unobscured power-law spectrum - a factor $\sim2$ on the final stacked flux and luminosity values - than for the Compton thick obscuration case, since the fraction of soft photons is much larger in the first case.

To estimate the spectral shape consistent with the stacking analysis results, we compared the detection in the UHB and the upper limits in the SB and MB with mock spectra for different obscuration values. The calculations used the physically motivated {\tt uxclumpy} code \citep{Buchner19} in XSPEC. The model specta cover a wide range of line-of-sight absorption column densities (log$N_H$=20--26 cm$^{-2}$). Two components describe the geometry of the clumpy obscuring gas. A toroidal distribution of clouds with variable vertical extent and column density parameterized by its covering factor (model parameter \texttt{TORsigma} in the range 0-0.84), plus an additional Compton thick (log $N_H$ = 25.5 cm$^{-2}$) reflector, which can be interpreted as part of the dust-free broad-line region, an inner rim, or a warped disk. The variable number of thick clouds parametrizes the covering factor of the ring  (model parameter \texttt{CTOR} in the range 0-0.6).
Given the very limited observational constraints, we fixed the slope of the primary continuum to $\Gamma=1.9$, and the high energy cut-off at $E_{cut}$ = 200 keV. We also assume a line of sight inclination of 60 degrees, consistent with the Type 2 classification of the sample. 

Model spectra were computed for a range of column densities in the high absorption regime (log$(N_H/\rm{cm}^{-2})>$ 23.5), and for a range of the \texttt{TORsigma} and \texttt{CTOR} parameters. In Figure~\ref{fig:spectra_uxclumpy_w_UL}, we report three model spectra representative of the various possible combinations of the parameters.

The red curve refers to a spectrum transmitted through a Compton thin ($N_H$ = 5$\times$10$^ {23}$ cm$^{-2}$) gas with high covering factor (0.84) and without an inner ring. The blue curve is for a Compton thick absorber ($N_H$ = 2$\times$10$^ {24}$ cm$^{-2}$) with the same assumptions for the covering factor and the inner ring. 
The green curve includes the inner Compton thick ring (\texttt{CTOR}=0.6) and is for a heavily obscured, $N_H$ = 10$^ {25}$ cm$^{-2}$, clumpy torus with high covering factor (0.84). 

For sources with such large line of sight column densities, changes in the covering factor of the clumpy clouds (i.e., in the \texttt{TORsigma} parameter) do not significantly affect the observed spectral shape in the energy range sampled in this work. 

The spectra are rescaled to the observed stacked flux in the 10--30 keV band. 
The upper limit in the 1--4 keV band is relatively loose, while the limit in the 4--7.25 keV band suggests that the average spectrum should steeply decline below the peak at $\sim$20 keV.
Such a behavior favors Compton thick absorption, almost independently of the details on the geometry of the clumpy absorber. Strong iron lines, with Equivalent Widths (EQW) of the order of 1--2 keV, are a distinctive feature of Compton thick absorption and are expected to contribute to the MB counts. In order to quantify their possible impact, we also tested the staking procedure in narrower bands centered at 6.4 keV, namely: 5.8-7.2 and 6.0-6.8 keV. 
There is no evidence for a signal with upper limits of about 20 and 10 counts, respectively. Given the \chandra energy resolution and the intrinsic weakness of our sources, narrower bands are severely photon starved. A putative iron K line with EQW of 2 keV at the median redshift ($z\approx4$) of our sample would contribute to $\sim20\%$ to the measured flux in the MB band. On the basis of the above-described test, we conclude that a strong iron line is fully consistent with our upper limits.


\begin{figure}
    \centering
    \includegraphics[width=0.48\textwidth,keepaspectratio]{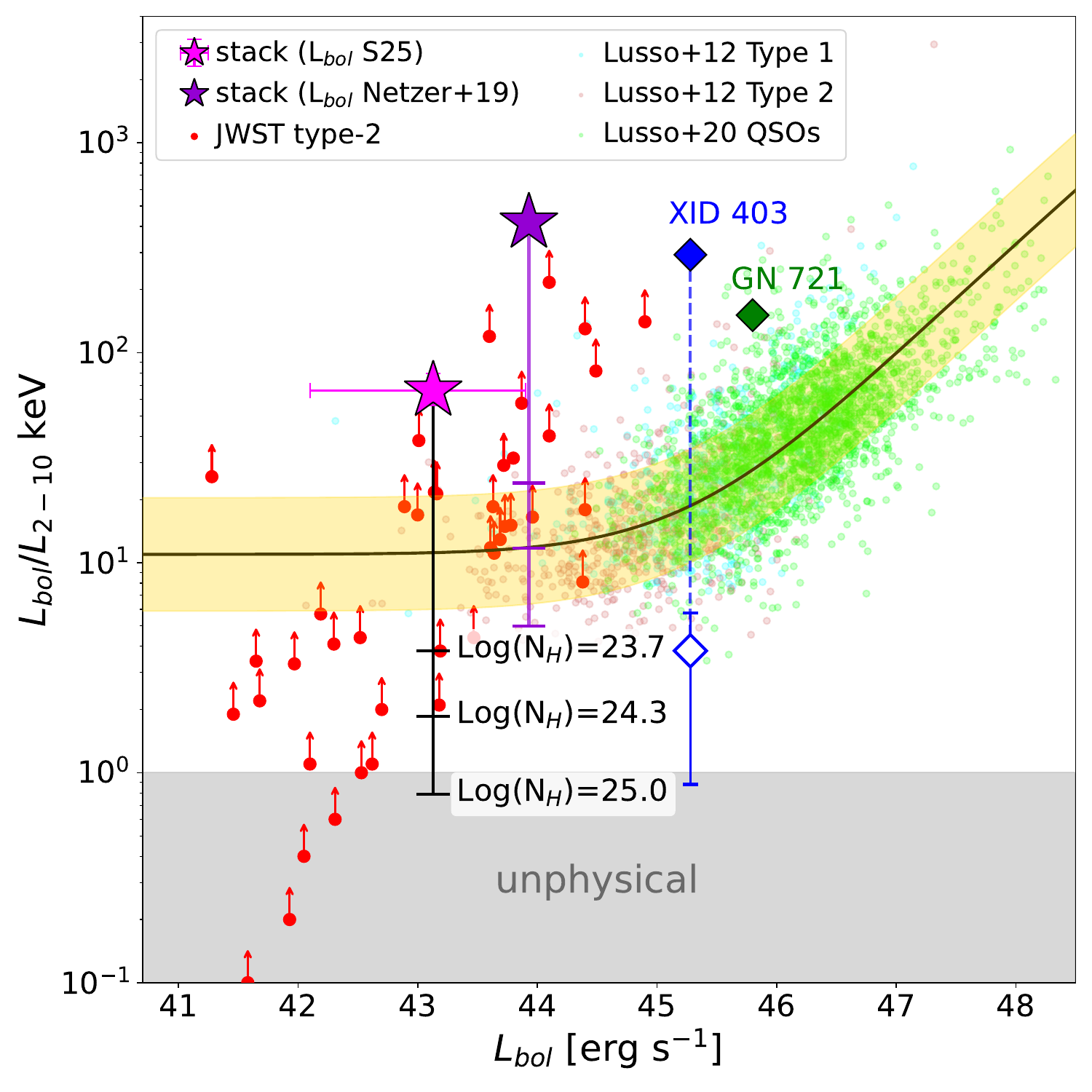}
    \caption{Ratio between the AGN bolometric luminosity and the X-ray luminosity as a function of the bolometric luminosity. Cyan and brown small points are X-ray selected Type 1 and Type 2 AGN in COSMOS (\citep{Lusso12}). Green small points are optically selected QSOs from \citep{Lusso2020}. The black continuous line shows the $k_{bol,X}-L_{bol}$ relation from \citep{Duras20} while the gold area shows the scatter.
    The red points are the lower limits obtained for each JWST Type 2 source in \citetalias{Maiolino25}.
    The magenta star shows the $k_{bol}$ obtained from the observed 2-10 keV luminosity derived from the 10-30 keV detection for a Compton thick spectrum with $N_H$ = $2\times10^{24}$ \cmsq (blue curve in Figure~\ref{fig:spectra_uxclumpy_w_UL}). The $k_{bol}$ corresponding to absorption-corrected intrinsic luminosities derived for  $N_H$ of $5\times10^{23}$,~$2\times10^{24}$ and $10^{25}$ \cmsq is shown with the black error bars. The violet data point is equivalent to the magenta one, but $L_{\rm bol}$ is derived from [O III] using the \cite{Netzer19} bolometric correction.
    The X-ray detected heavily obscured sources XID 403 and GN 721 are shown with blue and green diamonds, respectively. The $k_{bol,X}$ from de-absorbed luminosity is shown with the empty symbol for XID 403. 
    The region below $k_{bol}=1$ is marked as "unphysical" as it would imply that $L_{2-10keV}>L_{bol}$.
    }\label{fig:lbol_lx}
\end{figure}

\section{Discussion and Conclusions}\label{discussion}

The relatively tight upper limit in the MB coupled with the UHB detection indicates the obscured nature of JWST-selected Type 2 AGN.
The most straightforward interpretation is that they belong to the 
obscured population responsible for the X-ray background spectrum \citep{GCH07}.
In this respect, their nature is not exotic or elusive; they represent the low luminosity tail of the AGN population, which has already been seen and characterized at lower redshifts and higher luminosities. The bolometric to X-ray luminosity ratio is shown in Figure~\ref{fig:lbol_lx} for the Type 2 sample. The magenta star corresponds to the ratio computed assuming the observed 2-10 keV luminosity for Compton thick ($N_H = 2\times$10$^ {24}$ \cmsq) obscuration. This model is consistent with the upper limits in the SB and MB (blue curve in Figure~\ref{fig:spectra_uxclumpy_w_UL}). 
The bolometric correction corresponding to absorption-corrected luminosities for three different values of intervening absorption is also reported with thick black marks.  

The points are plotted at the median bolometric luminosity of the sample. Extreme values ($N_H \gtrsim 10^ {25}$ \cmsq) of absorption are ruled out as they would be inconsistent with the definition of the bolometric correction. Combined with the limits obtained by the X-ray stacking, the most probable average value for the absorption is of the order of log$N_H\sim24.2\pm0.3$ \cmsq.
This constraint is mitigated if a different recipe for the calculation of bolometric correction \citep{Netzer19} is adopted, as indicated by the 
violet star and the corresponding error bar.

We conclude that the average bolometric correction of the Type 2 AGN is consistent with the values already observed in the literature, reinforcing the hypothesis that the JWST-selected Type 2 AGN do not significantly differ from their lower redshift and/or higher luminosity counterparts.

For comparison, we show the location of the two X-ray detected and obscured sources in the parent sample: 
the blues diamond is XID 403 \citep{Gilli14}, where the  $k_{bol,X}$ is estimated from the observed (filled) and absorption-corrected (empty) luminosities from \citealt{Circosta19}.
The green diamond is GN 721, detected at $\sim 5\sigma$ with a limited counting statistic in the CDFN \citep{Maiolino25}. Its spectrum is flat ($\Gamma\sim 0.1$), consistent with Compton Thick obscuration. However, due to the limited number of detected photon counts, a proper spectral analysis is not possible, and therefore, the $k_{bol,X}$ from the de-absorbed luminosity cannot be derived.  
The two candidate Compton-thick broad-line AGN could represent the tip of the iceberg of a fainter and/or more obscured population.

A concern may arise from their abundance. If their space densities, at present ill-constrained, turn out to be much higher than those expected based on the extrapolation of the luminosity function at high redshift, then there might be tension with the observed intensity of the hard X--ray background.
However \cite{Grama}  showed that the current uncertainties on the level of the X-ray background around its peak at 20- 30 keV are such that it is possible to accommodate a relatively large number of heavily obscured Compton thick AGN with a spectrum similar to those reported in Figure~\ref{fig:spectra_uxclumpy_w_UL}.
A more detailed investigation on the space density of Type 2 AGN in various JWST surveys, including the effects of incompleteness, will be the subject of a future investigation. 

The lack of detection in any band for Type 1 confirms and extends previous claims about the puzzling nature of these objects. The upper limits in the rest-frame SB, MB, and UHB are shallower than those for the Type 2 sample due to the shorter total exposure (140 versus 209 Msec) and the fact that the majority of the Type 1 sample is in the CDF-N where the exposure is lower than in the CDF-S by a factor $\sim$ 3.5.
The available upper limits, computed separately for CDF-N and CDF-S, are not such to provide useful constraints on the shape of the average spectrum.  
A more tight constraint could be obtained by significantly increasing the sample size, especially in the CDF-S.  
However, it should be noted that, at high redshift, the softest energies probed are of the order of a few keV and thus the shape of the soft X-ray spectrum cannot be determined. For this reason we did not compute the median bolometric correction for the Type 1 sample. 
Future deep and wide X-ray surveys, such as those planned by the AXIS mission concept \citep{AXIS}, will probe the faint end of the luminosity function at high redshift \citep{Marchesi20} down to limiting fluxes one order of magnitude lower than \chandra deep surveys.

\begin{acknowledgements}

This research was supported by the Munich Institute for Astro-, Particle and BioPhysics (MIAPbP), which is funded by the Deutsche Forschungsgemeinschaft (DFG, German Research Foundation) under Germany´s Excellence Strategy – EXC-2094 – 390783311.
This research has made use of data obtained from the Chandra Data Archive, and software provided by the \chandra X-ray Center (CXC) in the application packages CIAO. 
\end{acknowledgements}

\bibliographystyle{aa}
\bibliography{biblio.bib} 

\appendix
\section{Properties of the sources used in this work}\label{sec:appendixA}
We report in this Appendix the list of sources analyzed in this work: in Table~\ref{Tab_phy_prop_type1} we report the sources classified as Type 1, while in Table~\ref{Tab_phy_prop_type2} we report the sources classified as Type 2. More details on the sample selection criteria are reported in Section~\ref{sample}.

\begin{table*}
	\caption{Type 1 AGN sample}
	\begin{tabular}{lccccc} 
		\hline
		\multicolumn{1}{l}{{ ID }} &
		\multicolumn{1}{c}{{ RA}} &
		\multicolumn{1}{c}{{ DEC }} &
		\multicolumn{1}{c}{{ $z$}} &
		\multicolumn{1}{c}{{ Reference}} \\
		\hline
		  159438  & 53.0544714 &  -27.902462    & 3.239 & \citetalias{Ignas25} \\ 
  9598    & 53.1618083  & -27.7707168  &    3.324 &  \citetalias{Ignas25} \\ 
  17341   & 53.0872694   & -27.7296227  &     3.598   & \citetalias{Ignas25} \\
  179198  & 53.0889771 &  -27.8606946 &    3.830  & \citetalias{Ignas25} \\
  13329   & 53.1390379  & -27.7844332  &    3.936  & \citetalias{Ignas25} \\
  172975  & 53.0877297  & -27.8712419  &   4.741 & \citetalias{Ignas25} \\
  8083    & 53.1328431  & -27.8018578  &    4.648 & \citetalias{Ignas25} \\ 
  38562   & 53.1358639  & -27.871645   &   4.822 & \citetalias{Ignas25} \\ 
  159717  & 53.0975283  & -27.9012603  &   5.077 & \citetalias{Ignas25} \\
  13971   & 53.1385932  & -27.7902534  &   5.480  & \citetalias{Ignas25}; 204851 in \citetalias{Maiolino25} \\
  3073    & 53.078875   & -27.884156   &   5.55 & \citetalias{Ignas25} \\
  10013704 & 53.1265350  & -27.8180923 &   5.918  & \citetalias{Ignas25} \\
30148179  & 53.142082   & -27.779848   &   5.921   & \citetalias{Ignas25}; GS9833 in \citetalias{Maiolino25} \\
  210600  & 53.1661147  &  -27.7720397 &   6.306 & \citetalias{Ignas25} \\
		\hline
   3608 & 189.117958  & 62.235528   &    5.269 &   \citetalias{Maiolino25} \\
  4014  & 189.300125  & 62.212028   &    5.228 & \citetalias{Maiolino25} \\
  9771  & 189.281000  & 62.247306   &    5.538 & \citetalias{Maiolino25} \\
 12839  & 189.344292  & 62.263361   &    5.241 & \citetalias{Maiolino25} \\
 13733  & 189.057083  & 62.268917   &    5.236 & \citetalias{Maiolino25} \\
 14409  & 189.072083  & 62.273417   &    5.139 & \citetalias{Maiolino25} \\ 
 15498  & 189.285542  & 62.280778   &    5.086 & \citetalias{Maiolino25} \\
 16813  & 189.179292  & 62.292528   &    5.355 & \citetalias{Maiolino25} \\
 z11    & 189.106083  & 62.242056   &   10.604 & \citetalias{Maiolino25} \\
   \hline
  LRD1 &  189.019240 &  62.243531  & 7.0388 & \cite{Meng25} \\
 LRD2  & 189.083488  &  62.202579  & 7.1883 & \cite{Meng25} \\
 \hline
 1085355  & 189.094365  &   62.198974  &  4.88 & \cite{Zhang25} \\
  1008671 & 189.161845    &    62.251054  & 4.41 & \cite{Zhang25} \\
  1029154 & 189.159025  &      62.260221 &  4.17 & \cite{Zhang25} \\
  1033320 & 189.125779    &    62.287404   & 4.48 &\cite{Zhang25} \\
  1034620 & 189.159764  &      62.295924   & 5.19  & \cite{Zhang25} \\
  1082263 & 189.212584    &    62.227436    & 3.98 & \cite{Zhang25} \\
  1086784 & 189.305706    &    62.236946    & 4.40  & \cite{Zhang25} \\
  1086855 & 189.286512    &    62.238138    & 4.41 & \cite{Zhang25} \\
  1087315 & 189.333584    &    62.246178    & 3.91 &  \cite{Zhang25} \\
  1089568 & 189.151821    &    62.272229    & 4.05 & \cite{Zhang25} \\
  1090549 & 189.235941    &    62.285544    & 5.20 & \cite{Zhang25} \\
  1008411 & 189.211089    &    62.250271    & 4.41 & \cite{Zhang25} \\
  \hline
  77652 & 189.293228  & 62.199003  & 5.229  & \citetalias{Ignas25} \\       
  73488 & 189.197396  & 62.177233  &  4.133      & \citetalias{Ignas25} \\  
  62309  & 189.248977  & 62.218350 &  5.172    & \citetalias{Ignas25} \\   
  61888 & 189.168016  & 62.217013  &   5.874     & \citetalias{Ignas25} \\   
  53757 & 189.269778  & 62.194208  & 4.447   & \citetalias{Ignas25} \\  
  38509 & 189.09144   & 62.22811   &  6.68    & \citetalias{Ignas25}  1146115 in \citetalias{Maiolino25} \\
  29648 & 189.209198  &  62.264268 &2.959     & \citetalias{Ignas25} \\  
  28074 & 189.064576  & 62.273820  & 2.259  & \citetalias{Ignas25} \\  
  20621 & 189.122515  & 62.292850  & 4.682   & \citetalias{Ignas25} \\  
  11836 & 189.220587  & 62.263675  & 4.409    & \citetalias{Ignas25} \\  
  2916  & 189.107739  & 62.269525  &   3.664      & \citetalias{Ignas25} \\  
  1093  & 189.179742  & 62.224629  &  5.594     & \citetalias{Ignas25} \\  
  954   & 189.151966  & 62.259635  &  6.759       & \citetalias{Ignas25} \\  
  \hline
	\end{tabular} \label{Tab_phy_prop_type1}\\
\end{table*}

\begin{table*}
	\caption{Type 2 AGN sample}
	\begin{tabular}{lcccccc} 
		\hline
		\multicolumn{1}{l}{{ ID }} &
		\multicolumn{1}{c}{{ RA}} &
		\multicolumn{1}{c}{{ DEC }} &
		\multicolumn{1}{c}{{ $z$}} &
		\multicolumn{1}{c}{{ log$(\frac{L_{bol}}{L_\odot})$}} &
		\multicolumn{1}{c}{{ Reference}} \\
		\hline
		GS 10040620 & 53.116542 & -27.762060  & 1.776 &   41.63 & \citetalias{Scholtz23}\\   
GS 8456  &   53.126083 & -27.800690 & 1.884  & 41.56 & \citetalias{Scholtz23}\\  
GS 209979  & 53.173542 & -27.773750 & 1.883  & 42.68 & \citetalias{Scholtz23}\\  
GS 10036017 & 53.105958 & -27.808470  & 2.016   & 44.08   & \citetalias{Scholtz23} \\
GS 10012511 & 53.120500 & -27.762750  & 2.019  & 41.66 & \citetalias{Scholtz23}\\   
GS 10008071 & 53.154458 & -27.771440  & 2.227  & 43.12  & \citetalias{Scholtz23} \\
GS 8880    & 53.127250 & -27.799310  & 2.327   & 41.95 & \citetalias{Scholtz23}\\  
GS 17670   & 53.125083 & -27.775720  & 2.350  & 42.99  & \citetalias{Scholtz23}\\
GS 10073   & 53.113333 & -27.795580  & 2.632  &  42.98 & \citetalias{Scholtz23}\\  
GS 10011849 & 53.120042 & -27.777860  & 2.686   & 43.58  & \citetalias{Scholtz23} \\
GS 7099    & 53.113792 & -27.805220  & 2.860  &  42.87 & \citetalias{Scholtz23}\\  
GS 114573  & 53.114750 & -27.795970  & 2.881   & 42.28   & \citetalias{Scholtz23}\\
GS 111511  & 53.154000 & -27.800920  & 3.008  &  42.50  & \citetalias{Scholtz23}\\ 
GS 132213  & 53.158542 & -27.772110  & 3.012   & 42.17   & \citetalias{Scholtz23}\\
GS 10013597 & 53.123250 & -27.769390 & 3.320 &  42.60  & \citetalias{Scholtz23}\\  
GS 10035295 & 53.114333 & -27.815500  & 3.588   & 43.70   & \citetalias{Scholtz23}\\
GS 104075  & 53.107042 & -27.812030 & 3.717  & 43.14 & \citetalias{Scholtz23}\\  
GS 108487  & 53.148875 & -27.805670  & 3.975  & 41.91  & \citetalias{Scholtz23} \\
GS 7762    & 53.113333 & -27.803000 & 4.146 &  43.62  & \citetalias{Scholtz23}\\ 
GS 95256   & 53.148958 & -27.822580  & 4.159  & 42.08  & \citetalias{Scholtz23} \\
GS 10000626 & 53.147000 & -27.813030 & 4.468 &  42.51 & \citetalias{Scholtz23} \\  
GS 111091  & 53.111667 & -27.801610  & 4.497  & 42.03  &  \citetalias{Scholtz23}\\
GS 17072   & 53.170208 & -27.777390  & 4.707  & 42.29  & \citetalias{Scholtz23}\\  
GS 10015338 & 53.115333 & -27.772890  & 5.073  & 43.85  & \citetalias{Scholtz23} \\
GS 4902    & 53.11852484 & -27.81296696	& 5.123 & 43.0 &  \citetalias{Scholtz23} \\
GS 9452    & 53.115833 & -27.797560  & 5.135  & 43.59  & \citetalias{Scholtz23}\\ 
GS 202208  & 53.164083 & -27.799720  & 5.450  & 44.88  & \citetalias{Scholtz23} \\
GS 208643$^{\bf a}$  & 53.130208 & -27.778360 & 5.566 & 43.61 & \citetalias{Scholtz23}\\ 
GS 16745$^{\bf a}$   & 53.130042 & -27.778389	&  5.574 & 43.7 & \citetalias{Scholtz23}\\  
GS 22251   & 53.154083 & -27.766080  & 5.804  & 44.08  & \citetalias{Scholtz23} \\
GS 10056849 & 53.113500 & -27.772830 & 5.820  & 43.16  & \citetalias{Scholtz23}\\ 
GS 201127   & 53.166833 & -27.804140  & 5.837   & 43.45  & \citetalias{Scholtz23} \\
GS 99671   & 53.126625 & -27.817720  & 5.936  & 43.17 & \citetalias{Scholtz23}\\  
GS 9422    & 53.121750 & -27.797640  & 5.942  & 44.47  &  \citetalias{Scholtz23}\\
GS 10013609 & 53.117292 & -27.764080 & 6.931  & 43.94 & \citetalias{Scholtz23} \\  
GS 10013905 & 53.118333 & -27.769000  & 7.206  &  43.67   & \citetalias{Scholtz23}\\
GS 21842   & 53.156833 & -27.767170 & 7.981  &  43.76  & \citetalias{Scholtz23}\\ 
GS 10058975 & 53.112417 & -27.774610  & 9.436   & 44.36  & \citetalias{Scholtz23} \\
		\hline
	\end{tabular} \label{Tab_phy_prop_type2}\\
	\tablefoot{$\rm ^a$ These two sources are at the same position (within the \chandra psf) and at very similar redshift and are therefore counted only once in the stack.  
    }
\end{table*}

\end{document}